\documentclass[aps,prd,showpacs,amsmath,amssymb,onecolumn,10pt]{revtex4}
\usepackage{graphicx}
\usepackage{amsmath}
\usepackage{amssymb}
\usepackage{amsfonts}
\usepackage{bm}
\usepackage{color}

\def\be{\begin{equation}}
\def\ee{\end{equation}}
\def\bea{\begin{eqnarray}}
\def\eea{\end{eqnarray}}

\newcommand{\f}[2]{\frac{#1}{#2}}

\begin{document}

\title{Weyl-Cartan-Weitzenb\"{o}ck gravity \\ as a generalization of
teleparallel gravity}
\author{Zahra Haghani$^1$}
\email{z_haghani@sbu.ac.ir}
\author{Tiberiu Harko$^2$}
\email{harko@hkucc.hku.hk}
\author{Hamid Reza Sepangi$^1$}
\email{hr-sepangi@sbu.ac.ir}
\author{Shahab Shahidi$^1$}
\email{s_shahidi@sbu.ac.ir}
\affiliation{$^1$Department of Physics, Shahid Beheshti University, G. C.,
Evin,Tehran 19839, Iran}
 \affiliation{$^2$Department of Physics and
Center for Theoretical and Computational Physics, The University
of Hong Kong, Pok Fu Lam Road, Hong Kong, P. R. China}
\date{\today}
\begin{abstract}
 We consider a gravitational model in a Weyl-Cartan space-time in which the
Weitzenb\"{o}ck condition of the vanishing of the sum of the curvature and
torsion scalar is imposed. In contrast to the standard teleparallel theories,
our model is formulated in a four-dimensional curved spacetime. The properties
of the gravitational field are then described by the torsion tensor and Weyl
vector fields. A kinetic term for the torsion is also included in the
gravitational action. The field equations of the model are obtained from a
Hilbert-Einstein type variational principle, and they lead to a complete
description of the gravitational field in terms of two fields, the Weyl vector
and the torsion, respectively,  defined in a curved background. The cosmological
applications of the model are investigated for a particular choice of the free
parameters in which the torsion vector is proportional to the Weyl vector. The
Newtonian limit of the model is also considered, and it is shown that the
Poisson equation can be
recovered in the weak field approximation. Depending on the numerical values of
the parameters of the cosmological model, a large variety of dynamic evolutions
can be obtained, ranging from inflationary/accelerated expansions to
non-inflationary behaviors. In particular we show that a de Sitter type late
time evolution can be naturally obtained from the field equations of the model.
Therefore the present model leads to the possibility of a purely geometrical
description of the dark energy, in which the late time acceleration of the
Universe is determined by the intrinsic geometry of the space-time.
\end{abstract}


\maketitle

\section{Introduction}

Despite its remarkable achievements at the scale of the Solar System and its
intrinsic theoretical beauty, general relativity, the best gravitational theory
proposed up to now, is facing presently some severe observational challenges.
The observations of high redshift supernovae \cite{1} and the
Boomerang/Maxima/WMAP data \cite{2}, showing that the location of the first
acoustic peak in the power spectrum of the microwave background radiation is
consistent with the inflationary prediction $\Omega = 1$, have provided
compelling evidence that about 95\% of the
content of the Universe resides in two unknown forms of matter and energy that
we call
cold dark matter (CDM) and dark energy (DE), with negative pressure,
respectively,  the first residing in bound objects as
non-luminous matter \cite{dm}, the latter in the form of a zero-point energy
that pervades the whole Universe \cite{PeRa03}. Dark matter is thought
to be composed of cold neutral weakly interacting massive particles, beyond
those existing in the Standard Model of Particle Physics, and not yet
detected in accelerators or in dedicated direct and indirect searches
\cite{OvWe04}. CDM contributes $\Omega _{DM} \approx 0.25$ to the total density
of the Universe, and is mainly
motivated by the theoretical interpretation of the galactic rotation curves and
large scale structure formation. DE  provides $\Omega _{DE}\approx 0.73$ to the
energy content of the Universe, and is responsible for the acceleration of the
distant type Ia supernovae \cite{PeRa03}.

There are two main theoretical lines of thought along which one could propose
explanations of the recent acceleration of the Universe. The first idea is that
the acceleration is produced by the presence of a ``real'' physical field in the
Universe, which may be also related to Einstein's cosmological constant. One way
of implementing this model
are cosmologies based on a mixture of cold dark matter and quintessence, a
slowly-varying, spatially
inhomogeneous component \cite{5}. An example of the idea of quintessence is the
suggestion that it
is the energy associated with a scalar field $Q$ with self-interaction potential
$V (Q)$. If the potential energy density is greater than the kinetic one, then
the pressure $p = \dot{Q}^2/ 2 - V (Q)$, associated to the $Q$-field is
negative. Quintessential
cosmological models have been intensively investigated in the physical
literature \cite{5}.
A different possibility has been followed in \cite{7}, where the conditions
under which the dynamics of a self-interacting Brans-Dicke (BD) field can
account for the accelerated expansion of the Universe have been analyzed.

Neither CDM nor DE have direct laboratory observational or experimental evidence
for their existence. Therefore
it would be important if a unified dark matter-dark energy scenario could be
found, in which these two components
are different manifestations of a single fluid. A candidate for such an
unification is the so-called generalized
Chaplygin gas, which is an exotic fluid with the pressure $p$ given as a
function of the density $\rho $ by the equation of state $p = -B/\rho ^n$, where
$B$ and $n$ are two parameters
to be determined \cite{8}. This dark energy-dark matter model was initially
suggested with $n = 1$, and then generalized for the case $n \neq 1$. The
cosmological implications of the Chaplygin gas model have also been considered
\cite{8}.

The second line of thought that can be considered for the explanation of the
recent observational data of the acceleration of the distant galaxies is based
on the assumption that the geometrical framework necessary to describe the
Universe must go beyond the Riemannian geometrical approach and the
Einstein-Hilbert Lagrangian on which standard general relativity is based. One
of the simplest possible theoretical extensions of general relativity are the
$f(R)$ gravity models, in which the Lagrangian of the gravitational field is
given by an arbitrary function of the Riemann curvature $R$. $f(R)$ models can
indeed explain the late acceleration of the universe, and their cosmological
implications have been intensively investigated \cite{9}.  The non-trivial
inclusion of the matter Lagrangian in the gravitational field Lagrangian has led
to the so-called $f\left(R,L_m\right)$ theories, which have been extensively
investigated recently \cite{10}.

An alternative geometrical approach, in which the basic physical variable is not
the metric $g_{\mu \nu}$ of the space-time, but a set of tetrad vectors
$e^i_{\;\;\mu }$ was recently considered in the so called $f(T)$ theories of
gravity, where $T$ is the torsion scalar. In this approach the torsion,
generated by the tetrad fields, can be used to describe general relativity
entirely using only torsion instead of curvature. This is the so-called
teleparallel equivalent of General Relativity (TEGR), which was introduced in
\cite{11}. Teleparallel gravity is based on the Weitzenb\"{o}ck space model
\cite{12}, in which torsion exactly compensates curvature and the space-time
becomes flat.  In the  $f(T)$
gravity models the field equations are of second order, unlike in $f(R)$
gravity, which in the metric approach is a fourth order theory. $f(T )$ models
have been extensively applied to cosmology, and in particular to explain the
late-time accelerating expansion without the need of dark energy \cite{13}.

Historically, the first extension of general relativity was proposed by Weyl
\cite{14}, who unified gravitation and electromagnetism by extending the notion
of parallel transport in general relativity to include the possibility that
lengths, and not only directions, change when vectors are transported along a
path. This non-integrability of length provides a geometrical interpretation for
the electromagnetic field and an elegant to unify the two known long-range
forces of nature. Despite its intrinsic beauty and rich physical structure,
Weyl's theory did not become a mainstream topic of research mainly because of
Einstein's comment \cite{15}  that ``...in Weyl's theory the frequency of
spectral lines would depend on the history of the atom, in complete
contradiction to known experimental facts.'' A generalization of Weyl's theory
was introduced  by Dirac \cite{16} who proposed the existence of an unmeasurable
metric $ds_E$, affected by transformations in the standards of length, and of a
measurable one, the conformally
invariant atomic metric $ds_A$.  In fact, any function $f(x)$ that transforms as
$f(x)/\sigma (x)$ under the transformation $g_{\mu \nu }\rightarrow \sigma ^2
g_{\mu \nu }$ would provide the appropriate relationship between the two metrics
as $f(x)ds_E = ds_A$.

The Weyl-Dirac theory and its applications were studied in \cite{17}, and some
of the early results are summarized in the book \cite{18}. The Weyl-Dirac theory
was considered within the framework of the weak field approximation in
\cite{19}, and it was shown that the resulting gravitational potential differs
from that of Newtonian one by a repulsive correction term increasing with
distance. If the time variation rate of gravitational coupling is adopted from
observational bounds, the theory can explain the rotation curves of typical
spiral galaxies without resorting to dark matter.  The intergalactic effects and
the gravitational lensing of clusters of galaxies have been estimated, and they
are consistent with observational data.

The Weyl geometry can be immediately generalized to include the torsion  of the
space-time \cite{He}. This geometric model is called the Weyl-Cartan geometry,
and it was extensively studied from both mathematical and physical point of view
\cite{20}. In \cite{21}  torsion was included in the geometric framework of the
Weyl-Dirac theory to build up an action integral, from which one can obtain a
gauge covariant (in the Weyl sense) general relativistic massive
electrodynamics. For a recent review of the geometric properties and of the
physical applications of the Riemann-Cartan and Weyl-Cartan space-times see
\cite{No}.

It is the purpose of the present paper to consider an extension of the
Weyl-Cartan gravitational model by explicitly including the Weitzenb\"{o}ck
condition that cancels torsion and curvature, in the geometric structure of the
Weyl-Cartan space. In this way we obtain a curvature-full geometric
gravitational model, in which the properties of the gravitational field are
determined by two fields, the Weyl vector and the torsion, as well as by some
scalars formed from the combination of these basic fields. The basic difference
with respect to the $f(T)$ type teleparallel theories is that the model is
formulated in the usual four-dimensional curved spacetime, whose properties are
described by the metric tensor  $g_{\mu \nu}$. However, the properties of the
gravitational field are described directly by two vector fields in a curved
space-time, and the metric is not the primary object determining the
gravitational properties. The field equations of this gravity model are obtained
from a variational principle, and
the cosmological applications of the field equations are studied in detail. In
particular we show that depending on the numerical values of the parameters of
the cosmological model, a large variety of dynamic evolutions can be obtained,
ranging from inflationary/accelerated expansions to non-inflationary behavior.
The weak field limit of the model is also considered, and it is shown that the
Poisson equation can be recovered from the Weitzenb\"{o}ck condition and the
field equations. This shows that the model proposed in the present paper is
phenomenologically viable.

The present paper is organized as follows. The geometrical properties of the
Weyl-Cartan space-time are briefly described in Section~\ref{sect2}. The
gravitational field equations corresponding to the Weyl-Cartan-Weitzenb\"{o}ck
(WCW) geometry are obtained in Section~\ref{sect3}. The Newtonian limit of the
model is also considered.  The cosmological implications of the model are
investigated in Section \ref{sect4}. Specific cosmological solutions are
presented in Section~\ref{sec5}. We discuss and conclude our results in
Section~\ref{sect5}. Finally, two appendices present the relation of this work
to phantom models and possible generalizations of the action, respectively.

\section{Geometrical preliminaries}\label{sect2}

Weyl generalized the Riemannian geometry of general relativity by supposing that
during the parallel transport around a closed path a vector would not only
undergo a change of direction, but would also experience a change in length
\cite{14}. In order to describe these two simultaneous changes mathematically,
Weyl introduced a vector field $w^{\mu }$, which, together with the metric
$g_{\mu \nu}$, represent the fundamental fields of the Weyl geometry. It is a
remarkable feature of the Weyl model that the properties of $w^{\mu }$ coincide
precisely with those of the electromagnetic potentials, suggesting that the
long-range forces of electromagnetism and gravity have a common geometric origin
\cite{16}.

If a vector of length $l$ is carried by parallel transport along an
infinitesimal displacement $\delta x^{\mu }$ in Weyl space, the change in its
length $\delta l$ is given by $\delta l=lw_{\mu }\delta x^{\mu }$ \cite{16}. For
parallel transport around a small closed loop of area $\delta s^{\mu \nu}$ the
change of the vector is $\delta l=lW_{\mu \nu}\delta s^{\mu \nu}$, where
\be
W_{\mu\nu}=\nabla_\nu w_{\mu}-\nabla_\mu w_{\nu},
\ee
where the covariant derivative is with respect to the metric.
Under the local scaling of lengths $\tilde{l}=\sigma (x)l$, the field $w_{\mu }$
transforms as $\tilde{w}_{\mu }=w_{\mu }+\left(\ln \sigma \right)_{,\mu }$,
while the metric tensor coefficients transform according to the conformal
transformations  $\tilde{g}_{\mu \nu }=\sigma ^2g_{\mu \nu}$ and $\tilde{g}^{\mu
\nu }=\sigma ^{-2}g^{\mu \nu}$, respectively \cite{18, No}. A distinctive featur
of the Weyl geometry is the presence of the semi-metric connection
\be\label{con}
\bar{\Gamma}^{\lambda}_{~\mu\nu}=\Gamma^{\lambda}_{~\mu\nu}+g_{\mu\nu}w^{\lambda
}-\delta^{\lambda}_{\mu}w_{\nu}-\delta^{\lambda}_{\nu}w_{\mu}
,
\ee
where $\Gamma^{\lambda}_{~\mu\nu}$ is the Christoffel symbol with respect to the
metric $g_{\mu\nu}$. In the Weyl geometry $ \bar{\Gamma}^{\lambda}_{~\mu\nu}$ is
assumed to be symmetric and with its help one can construct a gauge covariant
derivative \cite{No}. The Weyl curvature tensor can be written as
\be
\bar{R}_{\mu \nu \alpha \beta }=\bar{R}_{(\mu \nu )\alpha \beta }+\bar{R}_{[\mu
\nu ]\alpha \beta },
\ee
where we define
\begin{align}\label{eq}
\bar{R}_{[\mu \nu ]\alpha \beta }=R_{\mu\nu\alpha\beta}+2\nabla_\alpha
w_{[\mu}g_{\nu]\beta}+2\nabla_\beta w_{[\nu}g_{\mu]\alpha}+2w_\alpha
w_{[\mu}g_{\nu]\beta}+2w_\beta
w_{[\nu}g_{\mu]\alpha}-2w^2g_{\alpha[\mu}g_{\nu]\beta},
\end{align}
and
\be
\bar{R}_{(\mu \nu )\alpha \beta }=\frac{1}{2}\left(\bar{R}_{\mu \nu \alpha \beta
}+\bar{R}_{\nu \mu \alpha \beta }\right)=g_{\mu \nu}W_{\alpha \beta },
\ee
and square brackets denote anti-symmetrization. The first contraction of the
Weyl curvature  tensor is given by
\bea
\bar{R}_{~\nu }^{\mu }=\bar{R}_{\;\;\;\alpha \nu }^{\alpha \mu }=R_{~\nu }^{\mu
}+2w^{\mu }w_{\nu }+3\nabla _{\nu }w^{\mu }-\nabla _{\mu }w^{\nu }+
 g^{\mu }_{~\nu }\left(\nabla _{\alpha }w^{\alpha }-2w_{\alpha }w^{\alpha
}\right),
\eea
where $R^\mu_{~\nu}$ is the Ricci tensor constructed by the metric. For the Weyl
scalar we obtain
\be
\bar{R}=\bar{R}_{~\alpha }^{\alpha }=R+6\left(\nabla _{\mu }w^{\mu }-w_{\mu
}w^{\mu }\right).
\ee

The Weyl geometry can be extended to the more general case of the Weyl-Cartan
spaces with torsion, by considering a space time with a symmetric metric tensor
$g_{\mu \nu}$, defining the length of a vector,  and an asymmetric connection
$\hat{\Gamma }_{~\mu \nu}^{\lambda }$, which defines the law of the parallel
transport as $dv^{\mu }=-v^{\sigma } \hat{\Gamma }_{~\sigma \nu}^{\mu }dx^{\nu
}$ \cite{No, He}. In this case the connection may be decomposed into three
irreducible parts: the Christoffel symbol, the contortion tensor and the
non-metricity, so that one can write \cite{He}
\be\label{23}
\hat{\Gamma }_{~\mu \nu}^{\lambda }=\Gamma _{~\mu \nu}^{\lambda }+C_{~\mu
\nu}^{\lambda }+\frac{1}{2}g^{\lambda \sigma }\left(Q_{\nu \mu \sigma }+Q_{\mu
\nu \sigma}-Q_{\lambda \mu \nu }\right).
\ee

The contorsion $C_{~\mu \nu}^{\lambda }$ in equation ~(\ref{23}) can be given in
terms of the torsion tensor $\hat{\Gamma }_{~[\mu \nu ]}^{\lambda }$ defined as
\be
\hat{\Gamma }_{~[\mu \nu ]}^{\lambda }=\frac{1}{2}\left(\hat{\Gamma }_{~\mu \nu
}^{\lambda }-\hat{\Gamma }_{~\nu \mu }^{\lambda }\right),
\ee
as follows
\be
C_{~\mu \nu}^{\lambda }=\hat{\Gamma }_{~[\mu \nu ]}^{\lambda }+g^{\lambda \sigma
}g_{\mu \kappa }\hat{\Gamma }_{~[\nu \sigma ]}^{\kappa }+g^{\lambda \sigma
}g_{\nu \kappa }\hat{\Gamma }_{~[\mu \sigma ]}^{\kappa }.
\ee
The contorsion tensor is antisymmetric with respect to its first two indices.

The non-metricity tensor $Q_{\lambda \mu \nu }$ can be defined as (minus) the
covariant derivative of the metric tensor with respect to $\hat{\Gamma }_{~\mu
\nu}^{\lambda }$ \cite{He},
\be\label{con1}
Q_{\lambda \mu \nu}=-\frac{\partial g_{\mu \nu}}{\partial x^{\lambda }}+g_{\nu
\sigma }\hat{\Gamma }_{~\mu \lambda }^{\sigma }+g_{\sigma \mu }\hat{\Gamma
}_{~\nu \lambda }^{\sigma }
\ee

By comparing equations (\ref{con}) and (\ref{23}) it follows that the Weyl
geometry is a particular case of the Weyl-Cartan geometry, in which the torsion
is zero, and the non-metricity is given by $Q_{\lambda \mu \nu }=-2g_{\mu \nu
}w_{\lambda }$. Therefore in a Weyl-Cartan space  the connection is written as
\be\label{con3}
\hat{\Gamma}^{\lambda}_{~\mu\nu}=\Gamma^{\lambda}_{~\mu\nu}+g_{\mu\nu}w^{\lambda
}-\delta^{\lambda}_{\mu}w_{\nu}-\delta^{\lambda}_{\nu}w_{\mu}
+C^{\lambda}_{~\mu\nu},
\ee
where
\be
C^{\lambda}_{~\mu \nu}=T^{\lambda}_{~\mu \nu}-g^{\lambda \beta}g_{\sigma
\mu}T^{\sigma}_{~\beta\nu}
-g^{\lambda \beta}g_{\sigma \nu}T^{\sigma}_{~\beta\mu},
\ee
is the contortion of the Weyl-Cartan space, in which the Weyl-Cartan torsion
$T^{\lambda}_{~\mu \nu}$ is defined as
\be
T^{\lambda}_{~\mu \nu}=\frac{1}{2}\left(\hat{\Gamma}^{\lambda}_{~\mu
\nu}-\hat{\Gamma}^{\lambda}_{~\nu \mu}\right).
\ee
By the use of the connection, one can obtain the curvature tensor as
\be
K^{\lambda}_{~\mu\nu\sigma}=\hat{\Gamma}^{\lambda}_{~\mu\sigma,\nu}-\hat{\Gamma}
^{\lambda}_{~\mu\nu,\sigma}+\hat{\Gamma}^{\alpha}_{~\mu\sigma}\hat{\Gamma}^{
\lambda}_{~\alpha\nu}-\hat{\Gamma}^{\alpha}_{~\mu\nu}\hat{\Gamma}^{\lambda}_{
~\alpha\sigma}.
\ee
Using equation (\ref{con3}), one can obtain the curvature tensor
$K^{\lambda}_{~\mu\nu\sigma}$ in the terms of Riemann tensor and some new terms
containing Weyl vector, torsion and contortion. By contracting the resulting
curvature tensor, one can obtain the Ricci scalar of the model as follows
\begin{align}\label{eq}
K=K^{\mu\nu}_{~~\mu\nu}&=R+6\nabla_\nu w^\nu-4\nabla_\nu T^\nu-6w_\nu
w^\nu+8w_\nu T^\nu\nonumber\\
&+T^{\mu\alpha\nu}T_{\mu\alpha\nu}+2T^{\mu\alpha\nu}T_{\nu\alpha\mu}-4T_\nu
T^\nu.
\end{align}
where we define $T_\mu=T^\nu_{~\mu\nu}$ and all covariant derivatives are of the
metric.

%%%%%%%%%%%%%%%%%%%%%%%%%%%%%%%%%%%%%%%%%%%%%%%%%
%%%%%%%%%%%%%%%%%%%%%%%%%%%%%%%%%%%%%%%%%%%%%%%%%
%%%%%%%%%%%%%%%%%%%%%%%%%%%%%%%%%%%%%%%%%%%%%%%%%
\section{Field equations of the Weyl-Cartan-Weitzenb\"{o}ck
gravity}\label{sect3}

We work in a four dimensional Weyl-Cartan space-time with non-metricity and
torsion. In such a framework, at each point of space time, there is a symmetric
metric tensor $g_{\mu \nu}$, a Weyl vector $w_{\mu}$ and a torsion tensor
$T^{\lambda}_{~\mu\nu}$.
Using the units in which $16\pi G=1$, one may consider the following action
\begin{eqnarray}
I=\int d^4x \sqrt{-g}\bigg(K+\f{1}{4}W^{\mu \nu}W_{\mu \nu}+\beta \nabla_\mu T
\nabla^\mu T+L_m\bigg),
\end{eqnarray}
where $T=T_\mu T^\mu$ and
\be
W_{\mu\nu}=\nabla_\nu w_{\mu}-\nabla_\mu w_{\nu}.
\ee
The first term produces the kinetic term for the metric. The second term is the
usual kinetic term for the Weyl vector. We also add a kinetic term for the
torsion. We note that this is not the most general kinetic term for the torsion.
In Appendix \ref{apb} we will discuss an alternative possibility for the kinetic
term. In addition, we assume that the matter Lagrangian only depends on the
metric and not on the torsion and the Weyl vector.

In terms of the dynamical variables $\left(g_{\mu \nu }, w_{\mu
},T^{\lambda}_{~\mu\nu}\right)$ one can express the action  explicitly as
\begin{align}
I=\int
d^4x&\sqrt{-g}\bigg(R+T^{\mu\alpha\nu}T_{\mu\alpha\nu}+2T^{\mu\alpha\nu}T_{
\nu\alpha\mu}-4T_\mu T^\mu\nonumber\\
 &+\f{1}{4}W^{\mu \nu}W_{\mu \nu}+\beta \nabla_\mu T \nabla^\mu T-6w_\mu
w^\mu+8w_\mu T^\mu+L_m\bigg).
\end{align}

\subsection{ The Weitzenb\"{o}ck condition and the field equations of the
Weyl-Cartan-Weitzenb\"{o}ck gravity}\label{WCW}

Now, we assume the Weitzenb\"{o}ck condition
\be\label{tele}
R+T^{\mu\alpha\nu}T_{\mu\alpha\nu}+2T^{\mu\alpha\nu}T_{\nu\alpha\mu}-4T_\mu
T^\mu=0.
\ee
\\
Imposing this condition we obtain the following form for the action,
\begin{align}
I=\int d^4x \sqrt{-g}\bigg(\f{1}{4}W^{\mu \nu}W_{\mu \nu}+\beta \nabla_\mu T
\nabla^\mu T-6w_\mu w^\mu+8w_\mu T^\mu+L_m\bigg).
\end{align}
It is worth mentioning that, without the kinetic term for the torsion, the field
equations obtained by varying the action with respect to  the torsion will be
satisfied only when the Weyl vector vanishes which is the teleparallel gravity.
The field equations, obtained by variation of the action with respect to the
Weyl vector field and torsion are given by

\be\label{eqweyl1}
\f{1}{2}\nabla^\nu W_{\nu\mu}-6w_\mu + 4 T_\mu=0,
\ee
\bea\label{eqtorsion1}
2\left(w^\rho \delta^\sigma_\mu - w^\sigma \delta^\rho_\mu \right)
- \beta \left(T^\rho \delta^\sigma_\mu - T^\sigma \delta^\rho_\mu \right)\Box
T=0.
\eea
The variation of the action with respect to the metric gives us
\begin{align}\label{eqmetric1}
&\f{1}{2}\left(W_{\mu \rho}W_{\nu}^{\;\rho}-\frac{1}{4} g_{\mu\nu}
W_{\rho\sigma}W^{\rho\sigma}\right)-6\left(w_\mu w_\nu -
\frac{1}{2} g_{\mu\nu}w_\rho w^\rho\right)\nonumber\\
& + \beta \Big(\nabla_\mu T\nabla_\nu T -\frac{1}{2} g_{\mu\nu}\nabla_\rho
T\nabla^\rho T-2T_\mu T_\nu \Box T\Big)+4\left(T_\mu w_\nu +
T_\nu w_\mu -g_{\mu\nu}T_\rho w^\rho \right)
 -\f{1}{2}T^m_{\mu\nu}=0,
\end{align}
where $T^m_{\mu\nu}$ is the energy-momentum tensor.

The torsion tensor can be decomposed into its irreducible components under the
local $O(1,3)$ group as
\be\label{eqtors}
T_{\mu \nu \rho }=\f{2}{3}(t_{\mu\nu\rho}-t_{\mu\rho\nu})+\f{1}{3}\left(Q_\nu
g_{\mu\rho}-Q_\rho g_{\mu\nu}\right)+\epsilon_{\mu\nu\rho\sigma}S^\sigma,
\ee
where $Q_\mu$ and $S^\mu$ are two unknown vectors and $t_{\mu\nu\rho}$ is
symmetric with respect to the interchange of $\mu$ and $\nu$ and satisfies the
following identities
\begin{align}
t_{\mu\nu\rho}+t_{\nu\rho\mu}+t_{\rho\mu\nu}=0,\quad
g^{\mu\nu}t_{\mu\nu\rho}=0=g^{\mu\rho}t_{\mu\nu\rho}.
\end{align}
As one can see from equations \eqref{eqweyl1}, \eqref{eqtorsion1} and
\eqref{eqmetric1}, the torsion tensor appears in the contracted form, and so
only the $Q_\mu$ vector is relevant in the model. We can then consistently set
the tensor $t_{\mu\nu\rho}$ to zero and obtain the vector $S^\mu$ from the
Weitzenb\"{o}ck condition \eqref{tele}.
Contraction of  equation \eqref{eqtors} with $g^{\mu\rho}$
results in $T_\mu=Q_\mu$.  Now by contracting equation \eqref{eqtorsion1} with
$\delta^\mu_\sigma$ one obtains
\be
2w^\rho-\beta T^\rho\Box T=0.
\ee
Multiplication by $T_\rho$ and the assumption $T\neq 0$ leads to
\be\label{eqtor1}
\Box T=\f{2}{\beta}\f{1}{T}w^\rho T_\rho.
\ee
Now, putting this back to equation ~\eqref{eqtorsion1} one obtains
\be
T^\alpha
T_\alpha\left(w^\rho\delta^\sigma_\mu-w^\sigma\delta^\rho_\mu\right)=w^\alpha
T_\alpha\left(T^\rho\delta^\sigma_\mu-T^\sigma\delta^\rho_\mu\right),
\ee
with the solution
\be\label{eqT}
T_\mu=Q_\mu=Aw_\mu,
\ee
where $A$ is an arbitrary constant. The value of this constant can be obtained
by taking the covariant derivative of equation \eqref{eqweyl1} which leads to
\begin{align}
\nabla^\mu\left(T_\mu-\f{3}{2}w_\mu\right)=0,
\end{align}
So, we must have $A=3/2$.

Substituting this to equation\eqref{eqweyl1}, we obtain the following
differential equation for the Weyl vector
\be\label{eqWw}
\nabla^\nu\nabla_\mu w_\nu-\Box w_\mu=0.
\ee
This equation gives the Weyl vector as a function of the metric in general. With
a given matter energy-momentum tensor, one can then obtain the Weyl vector and
the metric components from equations \eqref{eqmetric1} and \eqref{eqWw}. The
vector $S^\mu$ however can be obtained from the teleparallel condition
\eqref{tele} as a function of the metric and  Weyl vector using  solution
\eqref{eqT}.

It is interesting to note that because our theory is generally covariant, we have $\nabla^\mu T^m_{\mu\nu}=0$ \cite{padmanabhan}. To see this, we
take the divergence of equation \eqref{eqmetric1} which leads to
\begin{align}
\nabla^\mu \bigg[&\f{1}{2}\left(W_{\mu \rho}W_{\nu}^{\;\rho}-\frac{1}{4} g_{\mu\nu} W_{\rho\sigma}W^{\rho\sigma}\right)-6\left(w_\mu w_\nu -
\frac{1}{2} g_{\mu\nu}w_\rho w^\rho\right)\nonumber\\
& + \beta \Big(\nabla_\mu T\nabla_\nu T -\frac{1}{2} g_{\mu\nu}\nabla_\rho T\nabla^\rho T-2T_\mu T_\nu \Box T\Big)+4\left(T_\mu w_\nu +
T_\nu w_\mu -g_{\mu\nu}T_\rho w^\rho \right)\bigg]=0.
\end{align}
This equation leads to nothing new since it is always zero by using the field equations. Note that the first parenthesis is the well-known
electromagnetic energy-momentum tensor and it is divergence free by using the first bianchi
identity and equation \eqref{eqWw}. The
remaining terms reduce to
\begin{align}
-3\nabla_\nu w^2+
\f{81\beta}{16}\big(\Box w^2\nabla_\nu w^2+\nabla^\mu w^2 \nabla_\mu \nabla_\nu
w^2-\nabla^\mu w^2 \nabla_\nu \nabla_\mu w^2\big)=0,
\end{align}
where $w^2=w^\mu w_\mu$. However, using equations \eqref{eqtor1} and \eqref{eqT} one can easily see that the LHS of the above equation gives zero.

\subsection{The Newtonian limit}\label{ap1}

Now we examine our theory in the Newtonian limit. Using equations (\ref{eqtors})
and (\ref{eqT}), one can write the Weitzenb\"ock condition equation
~(\ref{tele})  as
\begin{align}\label{eqWeit1}
R=6(w_\mu w^\mu-S_\mu S^\mu).
\end{align}
Taking the trace of equation ~(\ref{eqmetric1}) and using equations
(\ref{eqtor1}) and (\ref{eqT}) results in
\begin{align}\label{eqNew1}
-24w^2-\f{81\beta}{8} \nabla_\mu w^2 \nabla^\mu w^2 = T^m,
\end{align}
where $w^2=w_\mu w^\mu$ and $T^m$ is the trace of the matter energy-momentum
tensor. Assuming that matter can be described by a perfect fluid with energy
density $\rho $ and thermodynamic pressure $p$, after substituting equation
~\eqref{eqWeit1} into equation ~\eqref{eqNew1} one  obtains
\begin{align}\label{eqNew2}
R=\f{1}{4}(\rho-3p)-6S^2-\f{81\beta}{32}\nabla_\mu
w^2\nabla^\mu w^2,
\end{align}
where $S^2=S_\mu S^\mu$.

In the Newtonian limit, $R=-2R_{00}=-2\nabla^2\phi$, where $\phi$ is the
Newtonian potential corresponding to the $(00)$ component of the metric as
$g_{00}=-(1+2\phi)$. Also, as explicitly shown in the next Section, equation
\eqref{SS}, one can expect that  the norm of the vector $S_\mu$ must be
proportional to the norm of the Weyl vector. So we assume that in the Newtonian
limit the condition $S^2=bw^2$, $b={\rm constant}$, also holds. Consequently,
if we reintroduce  the factor $16\pi G$, we obtain the generalized
Poisson equation
\begin{align}\label{eqNew3}
\nabla^2\phi=-4\pi\f{1-b}{2} G (\rho-3p)+\f{81(1-b)}{64}\beta\nabla_\mu
w^2\nabla^\mu w^2,
\end{align}
which for $b=3$ takes the form of the generalized Poisson equation for weak
gravitational fields of the Weyl-Cartan-Weitzenb\"{o}ck gravity,
\begin{align}\label{eqNew4}
\nabla^2\phi=4\pi G (\rho-3p)-\f{81}{32}\beta\nabla_\mu w^2\nabla^\mu w^2.
\end{align}

If $\rho \gg p$ and for $(81/32)\beta\nabla_\mu w^2\nabla^\mu w^2\ll\rho $, we
obtain the standard Poisson equation of Newtonian gravity, $\nabla^2\phi=4\pi G
\rho $.

%%%%%%%%%%%%%%%%%%%%%%%%%%%%%%%%%%%%%%%%%%%%%%%%%
%%%%%%%%%%%%%%%%%%%%%%%%%%%%%%%%%%%%%%%%%%%%%%%%%
%%%%%%%%%%%%%%%%%%%%%%%%%%%%%%%%%%%%%%%%%%%%%%%%%
\section{Cosmological evolution equations in WCW gravity}\label{sect4}

To consider the cosmological solutions of the model we take the flat FRW metric
as
\be\label{metr}
ds^2=-dt^2+a(t)^2\left(dx^2+dy^2+dz^2\right),
\ee
and assume a perfect fluid form for the matter content of the universe.
>From the symmetry reasons shown by the metric above one may show that only the
time component of the Weyl vector is non-zero. We therefore write
\be\label{eqw}
w^\mu=\big[\psi(t),0,0,0\big],
\ee
where $\psi (t)$ is an arbitrary function of time.
Equation \eqref{eqT} implies that
\be\label{eqQ}
Q_\mu=-\f{3}{2}\big[\psi(t),0,0,0\big].
\ee
The argument above shows that with a given matter energy-momentum tensor, one
may obtain all the variables in  terms of the arbitrary temporal component of
the Weyl vector. As we have seen in section \ref{ap1}, one expects that for a
time-like ansatz for the Weyl vector, the vector $S^\mu$ must also be time-like.
 So, let's assume the following ansatz for $S^\mu$
\be\label{eqS}
S^\mu=\big[\Sigma(t),0,0,0\big],
\ee
where $\Sigma (t)$ is an arbitrary time dependent function.
The constraint equation \eqref{tele} gives the function $\Sigma(t)$ as
\be\label{SS}
\Sigma^2=\psi ^2+2H^2+\dot{H},
\ee
where $H=\dot{a}/a$ is the Hubble parameter.
Now,  equation  \eqref{eqmetric1} can be used to write
\begin{align}
 H&=\f{2}{\dot{\Psi}}\left[\f{1}{k}-\f{1}{6}\ddot{\Psi}\right],\label{eqH} \\
\rho &=-6\Psi +\f{k}{2} \dot{\Psi }^2,\label{eqrho} \\
p&=6\Psi +\f{k}{2} \dot{\Psi }^2, \label{eqp}
\end{align}
where we have denoted
\be
\Psi =\psi ^2,
\ee
and
\be
k=\frac{81\beta }{8},
\ee
respectively. Equation ~(\ref{eqH}) can be written as a second order linear
differential equation
\be\label{psisec}
\ddot{\Psi}+3H\dot{\Psi}-\frac{6}{k}=0,
\ee
or, equivalently
\begin{equation}
\frac{1}{a^{3}}\frac{d}{dt}\left( a^{3}\frac{d}{dt}\Psi \right) -\frac{6}{k}=0,
\end{equation}
giving the dependence of $\Psi $ on the Hubble function $H$ and of the scale
factor $a$. By taking the time derivative of equation ~(\ref{eqrho}) we obtain
\begin{equation}
\frac{d\rho}{dt}=-3kH\dot{\Psi}^{2},
\end{equation}
which proves the statement at the end of section \ref{WCW} on the conservation of
the energy momentum tensor in this setup
\be
\frac{d\rho}{dt}+3H\left(\rho+p\right)=0.
\ee
To go any further it would be useful to count the number of equations and
unknowns that one would have to deal with. A simple inspection shows that we
have three equations \eqref{eqH}, \eqref{eqrho} and \eqref{eqp} and four
unknowns, namely $H$, $\Psi$, $p$ and $\rho$.
In order to close the system of field equations we have to specify a
supplementary relation between $\rho$ and $p$. Once the equation of state is
given, the system of field equations (\ref{eqH})-(\ref{eqp}) is closed, and the
time evolution of the scale factor of the Universe and of the Weyl vector can be
obtained by assuming some appropriate initial conditions.

>From equations (\ref{eqH}) - (\ref{eqp}) it follows that for large varying $\Psi
$ the  pressure  acts like matter with a stiff equation of state, and as a
cosmological constant for small temporal variations of $\psi $
\begin{align}
p\approx \left\{
\begin{tabular}{ll}
$\rho= \frac{k}{2}\dot{\Psi}^2$, &$~~\dot{\Psi}^2\gg 12\Psi /k$, \\
$-\rho= 6\Psi$, & $~~\dot{\Psi}^2\ll 12\Psi /k$.
\end{tabular}\right.
\end{align}
Therefore, this result gives a geometrical interpretation of the dark energy in
terms of the Weyl vector. The equation of states are given by
\begin{align}
w=\frac{p}{\rho}\approx \left\{
\begin{tabular}{ll}
$1$, &$~~\dot{\Psi}^2\gg 12\Psi /k$, \\
$-1$, & $~~\dot{\Psi}^2\ll 12\Psi /k$.
\end{tabular}\right.
\end{align}

%%%%%%%%%%%%%%%%%%%%%%%%%%%%%%%%%%%%%%%%%%%%%%%%
%%%%%%%%%%%%%%%%%%%%%%%%%%%%%%%%%%%%%%%%%%%%%%%%
%%%%%%%%%%%%%%%%%%%%%%%%%%%%%%%%%%%%%%%%%%%%%%%%
\section{Cosmological models in WCW gravity}\label{sec5}

In this section we present a number of explicit, analytical and numerical
cosmological solutions of the field equations (\ref{eqH})-(\ref{eqp}) of the WCW
gravity model. As an indicator of the accelerating expansion we use the
deceleration parameter,  defined as
\begin{align}
q=\f{d}{dt}\f{1}{H}-1.
\end{align}
Positive values of $q$ indicate a decelerating behavior, while an accelerating
expansion of the Universe requires a negative $q$. In order to illuminate the
energy-momentum part of the geometric
content of the Universe, we write it as
\be
\rho(t)=\rho _m + \rho _T(t),
\ee
\be
p(t)=p_m+p_T(t),
\ee
where $\rho _m$ and $p_m$ represent the contribution of the energy density and
pressure of ordinary matter (dust and radiation) to the energy-momentum tensor,
and $\rho _T(t)$ and $p_T(t)$ are the effective energy-density and  pressure
corresponding to the Weyl vector and the torsion field.

\subsection{Cosmological models with a linear geometric equation of state}

As a first example of a cosmological model in WCW gravity we consider the case
in which the effective geometric pressure and density of the Weyl field satisfy
a linear, barotropic type equation of state of the form
\be
p _T=\left(\Gamma -1\right)\rho _T,
\ee
where $\Gamma $ is a constant. With this equation of state the gravitational
field equations become
\begin{widetext}
\begin{equation}
\frac{1}{a}\frac{da}{dt}=-\frac{4}{\sqrt{\big[2\left( \Gamma -2\right)
k\big]\big[%
\left( \Gamma -1\right) \rho _{m}-p_{m}+6\Gamma \Psi }\big]}\left\{ 1+\frac{%
\left( \Gamma -1\right) a\left( d\rho _{m}/da\right) -a\left(
dp_{m}/da\right) }{6\left[ \left( \Gamma -1\right) \rho _{m}-p_{m}+6\Gamma
\Psi \right] }\right\} ^{-1}, \label{s1}
\end{equation}
and
\begin{equation}\label{s2}
\frac{d\Psi }{dt}=\sqrt{\frac{2}{k\left( \Gamma -2\right) }\bigg(\left(
\Gamma -1\right) \rho _{m}-p_{m}+6\Gamma \Psi \bigg)}.
\end{equation}
\end{widetext}
In obtaining the above solution we have assumed that $\dot{\Psi}\geq0$. The
system of equations (\ref{s1}) and (\ref{s2}) must be supplemented
with the scale factor dependence on $\rho _{m}$ and $p_{m}$, which can be
obtained from the ordinary matter equation of state $p_{m}=p_{m}\left( \rho
_{m}\right) $ and from the energy conservation equation, and must be
integrated with some appropriate initial conditions $a\left( 0\right) =a_{0}$
and $\psi \left( 0\right) =\psi _{0}$, respectively. In the following we
consider that the matter content of the Universe consists of pressure-less dust,
with $p_m=0$. Hence the energy density of the matter varies as $\rho _m=\rho
_0/a^3$, where $\rho _0$ is the present day matter density. By introducing a
dimensionless time $\tau $, and the dimensionless Weyl vector $\theta $, defined
as
\begin{equation}
\tau =6\Gamma \sqrt{\frac{2}{k\left( \Gamma -1\right) \left( \Gamma
-2\right) \rho _{0}}}t,
\end{equation}
and
\begin{equation}
\theta =\frac{6\Gamma }{\left( \Gamma -1\right) \rho _{0}}\Psi ,
\end{equation}
respectively, the cosmological field equations of the WCW gravity model with
a linear geometric equation of state take the form
\begin{equation}
\frac{da}{d\tau }=-\frac{2}{3\Gamma }\frac{\sqrt{1+a^{3}\theta }}{%
1+2a^{3}\theta }a^{5/2},
\end{equation}
\begin{equation}
\frac{d\theta }{d\tau }=\sqrt{\frac{1}{a^{3}}+\theta }.
\end{equation}
One can solve the above equations analytically for $\theta(\tau)$ as a function
of $a(\tau)$,
\begin{align}\label{eqtheta}
\theta (\tau)=-\f{\Gamma}{2(\Gamma-1)}a^{-3}+C_1a^{-3\Gamma},
\end{align}
where $C_1$ is an integration constant. The solution for the scale factor can
then be obtained implicitly as a function of time
\begin{align}\label{eqaaa}
\tau
=\f{3\Gamma}{\sqrt{2(\Gamma-1)}}\int\f{a^{-3\Gamma-\f{5}{2}}\left[a^{3\Gamma}
-2C_1(\Gamma-1)a^3\right]}{\sqrt{2C_1(\Gamma-1)a^{3-3\Gamma}+\Gamma-2}}\textmd{d
}a.
\end{align}

\begin{figure}
 \centering
 \includegraphics[scale=0.80]{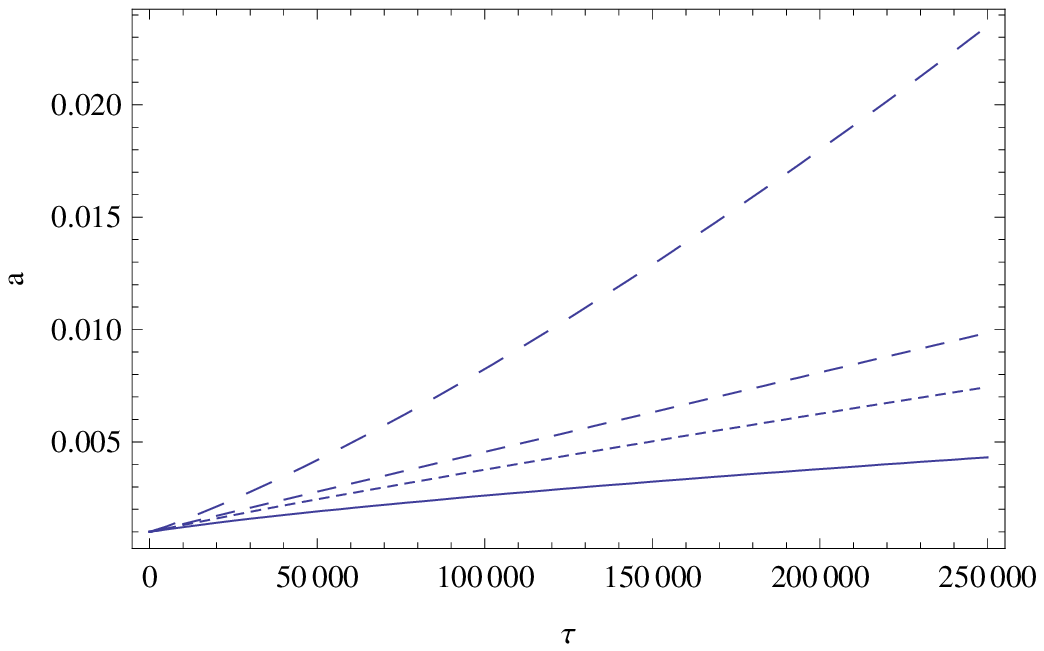}
 \caption{Time variation of the scale factor of the Universe in the WCW gravity
model with linear barotropic equation of state for different values of $\Gamma
$: $\Gamma =-1$ (solid curve), $\Gamma =-3/4$ (dotted curve), $\Gamma =-2/3$
(dashed curve) and $\Gamma =-1/2$ (long dashed curve), respectively. The initial
conditions for the cosmological evolution are $a(0)=0.001$ and $\theta
(0)=10^{-6}$. }
 \label{scale}
\end{figure}

\begin{figure}
 \centering
 \includegraphics[scale=0.80]{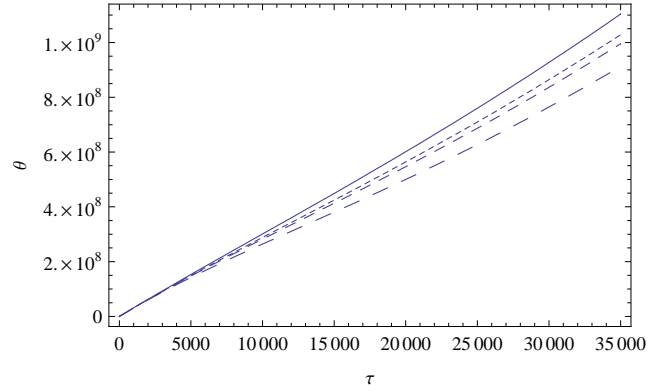}
 \caption{Time variation of the Weyl vector in the WCW gravity model with linear
barotropic equation of state for different values of $\Gamma $: $\Gamma =-1$
(solid curve), $\Gamma =-3/4$ (dotted curve), $\Gamma =-2/3$ (dashed curve) and
$\Gamma =-1/2$ (long dashed curve), respectively. The initial conditions for the
cosmological evolution are $a(0)=0.001$ and $\theta (0)=10^{-6}$. }
 \label{theta}
\end{figure}

The time variations of the scale factor $a$ and of the dimensionless Weyl vector
$\theta $ are presented, for different values of $\Gamma $, in Figs.~\ref{scale}
and \ref{theta}, respectively.

\begin{figure}
 \centering
 \includegraphics[scale=0.80]{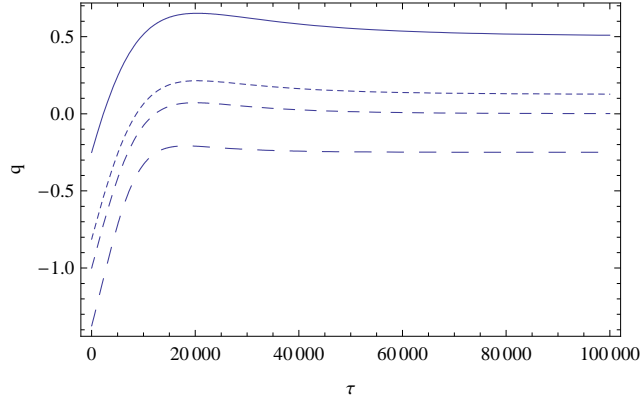}
 \caption{Time variation of the deceleration parameter in the WCW gravity model
with linear barotropic equation of state for different values of $\Gamma $:
$\Gamma =-1$ (solid curve), $\Gamma =-3/4$ (dotted curve), $\Gamma =-2/3$
(dashed curve) and $\Gamma =-1/2$ (long dashed curve), respectively. The initial
conditions for the cosmological evolution are $a(0)=0.001$ and $\theta
(0)=10^{-6}$. }
 \label{q_bar}
\end{figure}

The time variation of the deceleration parameter for this class of cosmological
models is presented in Fig.~\ref{q_bar}. As one can see from the figure,
depending on the value of $\Gamma $, a large variety of cosmological behaviors
can be obtained. Here, the universe may start from an accelerating state, with
negative $q$, and depending on the value of $\Gamma $,  can end either in a
decelerating phase, or with a constant negative value of the deceleration
parameter. This class of models may be relevant for the description of the
inflationary epoch of the very early evolution of the Universe. For $\Gamma
<-1$, the cosmological models are decelerating with $q>0$ for all times.
In the absence of matter $\rho _m=p_m=0$ and for $\psi\geq0$,  equations
(\ref{s1}) and (\ref{s2}) have the simple solution
\begin{align}
\psi (t)=\sqrt{\frac{3\Gamma }{\left(\Gamma -2\right)k}}~t,\; a(t)=a_0t^n,
\end{align}
where
\be
n=-\frac{2}{3\Gamma }.
\ee
In order to have an expansionary evolution it is necessary that $\Gamma <0$. The
deceleration parameter for this model is
\be
q=\frac{1}{n}-1=-\frac{3\Gamma }{2}-1.
\ee
The Universe will experience an accelerated, power law inflationary expansion if
the condition $\Gamma >-2/3$ is satisfied.

\subsection{Exponentially accelerating solutions}

 In the following we consider first the case of a vacuum Universe with $\rho
_m=p_m=0$. From equation ~(\ref{eqH}) it immediately follows that $H=$ constant
if $ \dot{\Psi}=\psi _0^2=$constant$\neq 0$, giving
 \be
 \psi (t)=\psi _0\sqrt{t}.
 \ee
 For the geometric density and pressure we obtain
 \be
 \rho_T=\psi _0^2\left(\frac{k}{2}\psi _0^2-6t\right),\;p_T=\psi
_0^2\left(\frac{k}{2}\psi _0^2+6t\right).
 \ee
 The expansion of the Universe is accelerating with the scale factor given by
$a=a_0\exp\left(H_0t\right)$ where the Hubble constant is $H_0=2/k\psi _0^2$. In
order to find the general conditions for a de Sitter type expansion we start
from equation ~(\ref{psisec}), written as
 \begin{equation}
\frac{d^{2}}{dt^{2}}\Psi +3H\frac{d}{dt}\Psi -\frac{6}{k}=0.
\label{eqpsi}
\end{equation}

For an accelerated expansion $H=H_{0}=$ constant, and therefore the general
solution of equation ~(\ref{eqpsi}) is given by
\be\label{psiexp}
 \psi (t)=\sqrt{\frac{2 }{H_0 k}t-\frac{ C_1}{3
   H_0}e^{-3 H_0 t}+C_2},
 \ee
 where $C_1$ and $C_2$ are arbitrary constants of integration and we assume that
$\psi\geq0$. The energy density and pressure of the Universe are given by
 \bea
 \rho _T=-6C_2 + \frac{4C_1}
   {H_0}e^{-3H_0t} +
  \frac{{C_1}^2k}
   {2}e^{-6H_0t} +\frac{2\left(1 - 6H_0t\right)\texttt{}}{{H_0}^2k},
 \eea
 and
 \be
 p_T=6C_2 + \frac{C_1^2k}
   {2}e^{-6H_0t} +
  \frac{2\left(1 + 6H_0t\right)}{{H_0}^2k},
 \ee
 respectively. At late times, the energy density and the pressure satisfy the
relation $\rho _T+p_T=4/H_0^2k$.

\subsection{Power law expansion}

\begin{figure}
 \centering
 \includegraphics[scale=0.55]{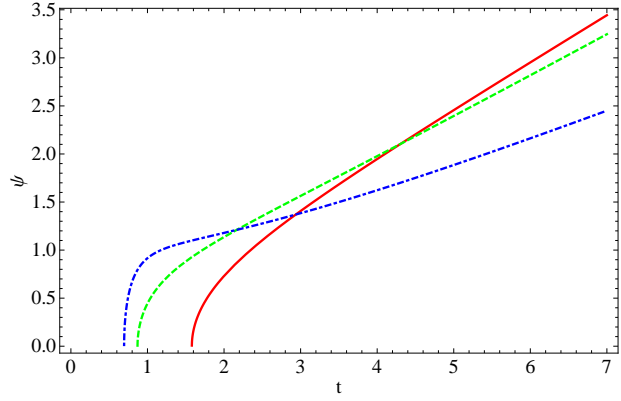}
 \caption{Time evolution of the temporal component of the Weyl vector for
$C_1=1$, $C_2=1$, $\beta=1/4$ and $\lambda=1/2$ (solid curve), $\lambda =2/3 $
(dashed curve), and $\lambda =8/5$ (dot-dashed curve).}
 \label{fig2}
\end{figure}

Let us assume that the scale factor behaves as $a(t)=a_0t^{\lambda}$. The
temporal component of the Weyl vector can be obtained from  equation
\eqref{eqpsi} with the result
\be\label{psipow}
\psi(t)=\sqrt{\f{3}{k(3\lambda+1)}t^2+\f{C_1}{1-3\lambda}t^{1-3\lambda}+C_2},
\ee
where $C_1$ and $C_2$ are integration constants and we assume that $\psi\geq0$.
In this case, the energy-density and pressure are given by
\bea
\rho_T=-\f{54\lambda}{k(1+3\lambda)^2}t^2-\f{36C_1\lambda}{1-9\lambda^2}t^{
1-3\lambda}+\f{kC_1^2}{2}t^{-6\lambda}-6C_2,
\eea
and
\bea
p_T=-\f{18(2+3\lambda)}{k(1+3\lambda)^2}t^2+\f{12C_1}{1-9\lambda^2}t^{1-3\lambda
}+\f{kC_1^2}{2}t^{-6\lambda}+6C_2,
\eea
respectively.
Fig.~\ref{fig2} shows the behavior of the Weyl vector for three cases of matter
dominated, radiation dominated and the late time acceleration phase of the
universe with $\lambda=2/3,1/2$ and $8/5$, respectively.

\begin{figure}
 \centering
 \includegraphics[scale=0.6]{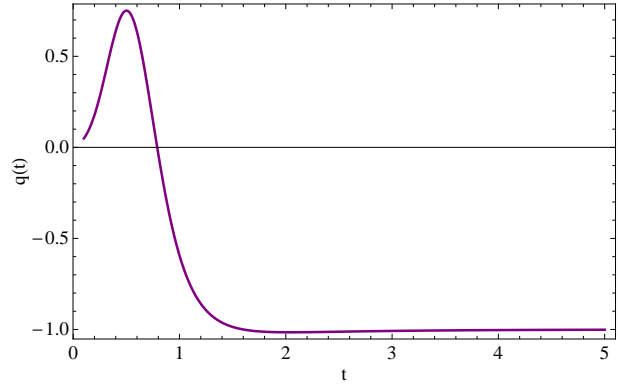}
 \caption{Deceleration parameter as a function of time for $\psi_1=0.1$,
$\psi_2=10$, $\psi _3=-1$, $\psi_4=-1$ and $\beta=1/4$.}
 \label{fig3}
\end{figure}

As can be seen form the above solutions one may construct a model starting from
a matter dominated universe and ending with a self-accelerating universe by
choosing appropriate values for $\psi (t)$. In order to do this one must choose
a function which behaves like \eqref{psipow} for small $t$, and behaves like
\eqref{psiexp} for large $t$.
Let us assume that
\begin{align}
\psi(t)^2=\psi_1 t^{\f{6}{5}}+\psi_2 t+\f{\psi_3}{t}+\f{\psi_4}{t^2},
\end{align}
with $\psi _i$ some constants.
In Fig.~\ref{fig3} we have plotted the deceleration parameter as a function of
time.
As can be seen from the figure, the universe accelerates at late times with the
deceleration parameter $q\approx -1$, and decelerates in the early epoch, in
agreement with observational data.

\section{Discussions and final remarks}\label{sect5}

In this paper we have investigated a geometrical gravitational theory based on
the imposition of the Weitzenb\"{o}ck condition, a relation between the
curvature and torsion, in a Weyl-Cartan space-time.  As a result, the
Einstein-Hilbert action of the gravitational field becomes independent of the
curvature, thus leading to the theoretical possibility of the description of the
gravitational phenomena in terms of two interacting fields, the Weyl vector and
torsion, in a curved background geometry, characterized by a symmetric metric
tensor. A dynamical torsion kinetic term has also been added to the
gravitational action. By performing the independent variation of the action with
respect to the vector fields and metric, the field equations are given by a set
of three independent second order partial differential equations, describing the
dynamics of the two vector fields in the background geometry, and in the
presence of the standard matter fields. The model does not contain an explicit
dynamical equation for
the metric, which has to be obtained generally by solving the Weitzenb\"{o}ck
constraint equation.

The weak field limit of the model has also been investigated, and it was shown
that the Poisson equation can be recovered from the  Weitzenb\"{o}ck condition
and the field equations. The existence of a Newtonian/weak field limit shows
that the present model is phenomenologically viable.

The cosmological implications of the model have been analyzed by assuming a flat
Friedmann-Robertson-Walker type metric. The resulting field equations can be
reduced to three independent equations, giving the Hubble function and the total
energy density and pressure as a function of the Weyl vector and its time
derivative only.

Present day observations do not rule out the possibility of the presence of some
extra gravitational effects, beyond the standard general relativity model,
acting at the Solar System, galactic and cosmological levels.
The predictions of the WCW gravity model could lead to some new effects, as
compared to the
predictions of general relativity, or other generalized gravity models, in
several problems of current interest such as cosmology, dark matter,
gravitational collapse or the generation of gravitational waves. The study of
these phenomenon may also provide some specific signatures and effects which
could distinguish and discriminate between  various gravitational models. In
order to explore in more detail the connections between the WCW gravity model
and the properties of standard gravity at different length scales, some explicit
physical models are necessary to be built. In particular the properties of the
static, spherically symmetric gravitational fields in vacuum and inside stars
should be considered and predictions of the model should be compared with the
existing observational data at the Solar System level. These studies will be
pursued  in the forthcoming works.

\acknowledgments

We would like to thank to the anonymous referee for comments and suggestions that helped us to significantly improve our manuscript.

\appendix
\section{Relation to Phantom model}
An interesting result of the model is the form of the geometric equation of
state in the absence of matter, $\rho _m=p_m=0$, given by equations
(\ref{eqrho}) and (\ref{eqp}), respectively, as well as the evolution equation
for the Weyl field, given by equation ~(\ref{psisec}). By rescaling the Weyl
field as $\Psi\rightarrow \phi /\sqrt{k}$, $k>0$,  the density and the pressure
of the Weyl field take a form similar to the energy and pressure of the standard
scalar field model, with $\rho _T=\dot{\phi}^2/2-V(\phi)$, and
$p_T=\dot{\phi}^2/2+V(\phi)$, respectively, with $V\left(\phi
\right)=(6/\sqrt{k})\phi$. However, as compared to the case of the standard
scalar field models, the potential  has the opposite sign. Moreover, in the
Klein - Gordon equation $\ddot{\phi}+3H\dot{\phi}-dV/d\phi =0$, describing the
evolution of the Weyl field, the term $dV/d\phi $ appears again with the wrong
sign as compared to the case of the standard Klein - Gordon equation
\be\label{KG}
\ddot{\phi}+3H\dot{\phi}+dV/d\phi =0.
\ee
 Fields satisfying equation ~(\ref{psisec}) instead of equation ~(\ref{KG}) are
known as phantom fields \cite{phan}. In a cosmological context the hypothetical
phantom scalar field would cause super-acceleration of the universe, with the
Hubble parameter increasing with time, $\dot{H} >0$,  and leading  to a Big Rip
singularity at a finite time in the future. Hence in the present model the Weyl
field behaves like an equivalent phantom scalar field \cite{phan}, and
super-accelerating cosmological models can indeed be explicitly obtained.

 The case of scalar fields with ``wrong'' sign in the Klein - Gordon equation
was discussed recently in \cite{Fa}, in the context of the analysis of the
correspondence between scalar fields and effective perfect fluids. In the
framework of the scalar field description of fluid systems, the Klein - Gordon
equation with the ``wrong'' sign can be obtained from a Lagrangian of the form
$L_{\rho _{\phi }}=-a^3\rho _{\phi }=-a^3\left[\dot{\phi }^2/2+V(\phi )\right]$,
while the ``correct'' sign in the  Klein - Gordon equation is obtained for
$L_{p_{\phi }}=a^3p_{\phi }=a^3\left[\dot{\phi }^2/2-V(\phi )\right]$. In the
present model a similar effect of the sign change in the Klein-Gordon equation
is related to the change in the sign of the potential in the geometric density
and pressure associated to the Weyl field.
\section{generalizing the action}\label{apb}
One can add another type of the kinetic term for the torsion to the action,
namely, $\alpha T_{\mu\nu}T^{\mu\nu}$, $\alpha ={\rm constant}$. Another
interesting term which can be added to the action is an interaction term between
torsion and the Weyl vector, $\gamma T^\alpha_{~\mu \nu}\nabla_\alpha W^{\mu
\nu}$, $\gamma ={\rm constant}$, which was originally introduced by Israelit
\cite{18, 21}. With this additional terms the equations of motion can be
generalized to
\be\label{eqweyl}
\f{1}{2}\nabla^\nu W_{\nu\mu}-6w_\mu + 4 T_\mu+\gamma \nabla^\nu \nabla_\alpha
T^\alpha_{\;\mu\nu}=0,
\ee
\bea\label{eqtorsion}
4\left(w^\rho \delta^\sigma_\mu - w^\sigma \delta^\rho_\mu \right)+2\alpha
\left( \delta^\sigma_\mu \nabla_\alpha T^{\rho \alpha} -\delta^\rho_\mu
\nabla_\alpha T^{\sigma\alpha} \right)- 2\beta \left(T^\rho \delta^\sigma_\mu -
T^\sigma \delta^\rho_\mu \right)\Box T+\gamma\nabla_\mu W^{\rho\sigma}=0,
\eea
and
\begin{align}\label{eqmetric}
&\f{1}{4}\left(2W_{\mu \rho}W_{\nu}^{\;\rho}-\frac{1}{2} g_{\mu\nu}
W_{\rho\sigma}W^{\rho\sigma}\right)-6\left(w_\mu w_\nu - \frac{1}{2}
g_{\mu\nu}w_\rho w^\rho\right)+\alpha (2T_{\mu\rho}T_\nu^{\;\rho } -\frac{1}{2}
g_{\mu\nu} T_{\rho\sigma}T^{\rho\sigma})\nonumber\\
& -\gamma\Bigg[\frac{1}{2}g_{\mu\nu}\nabla_\gamma
W^{\rho\sigma}T^{\gamma}_{\;\;\rho\sigma}-\nabla_\rho T^{\rho\sigma}_{\;\;\;\nu}
W_{\mu \sigma}
-\nabla_\rho T^{\rho\sigma}_{\;\;\;\mu} W_{\nu \sigma}+
\frac{1}{2}\nabla_\rho \Big( T_\nu^{~\rho\sigma}W_{\mu\sigma}-
 T_{\nu\mu}^{\;\;\;\sigma}W^\rho_{\;\sigma}+T_{\mu}^{\;\;\rho\sigma}W_{\nu\sigma
}-T_{\mu\nu}^{\;\;\;\sigma} W^\rho_{\;\sigma}\Big)\Bigg] \nonumber\\
&+ \beta \Big(\nabla_\mu T\nabla_\nu T -\frac{1}{2} g_{\mu\nu}\nabla_\rho
T\nabla^\rho T-2T_\mu T_\nu \Box T\Big)+4\left(T_\mu w_\nu + T_\nu w_\mu
-g_{\mu\nu}T_\rho w^\rho \right)-\f{1}{2}T^{m}_{~\mu\nu}
=0,
\end{align}
respectively. The effects of such terms as above on the theory would require
further investigations
which are ongoing.

\end{document}